\documentclass{article}

\usepackage[preprint]{neurips_2024}

\usepackage[utf8]{inputenc} 
\usepackage[T1]{fontenc}    
\usepackage{url}            
\usepackage{booktabs}       
\usepackage{amsfonts}       
\usepackage{amsmath}
\usepackage{bm}
\usepackage{nicefrac}       
\usepackage{microtype}      
\usepackage{xcolor}         
\usepackage{graphicx}
\usepackage{layouts}
\usepackage{subcaption}
\usepackage{verbatim}
\usepackage{todonotes}
\usepackage{makecell}
\usepackage{soul}
\usepackage{longtable}
\usepackage{listings}
\usepackage[breakable]{tcolorbox}
\usepackage{siunitx}
\usepackage{hyperref}       

\renewcommand{\vec}[1]{\boldsymbol{#1}}

\title{How Personality Traits Shape LLM Risk-Taking Behaviour}

\author{
    John Hartley$^{1,*}$ \quad
    Conor Hamill$^1$ \quad
    Devesh Batra$^1$ \quad
    Dale Seddon$^1$ \vspace{0.1em}\\
    \textbf{Ramin Okhrati}$^2$ \quad
    \textbf{Raad Khraishi}$^{1,2}$ \vspace{0.3em}\\
    $^1$NatWest AI Research\\
    $^2$University College London\\
}

\begin{document}
\maketitle
\renewcommand{\thefootnote}{\fnsymbol{footnote}}
\footnotetext[1]{Corresponding author: john.hartley@natwest.com}
\renewcommand{\thefootnote}{\arabic{footnote}}

\begin{abstract}
Large Language Models (LLMs) are increasingly deployed as autonomous agents,
necessitating a deeper understanding of their decision-making behaviour under risk.
This study investigates the relationship between LLMs' personality traits and risk propensity,
employing cumulative prospect theory (CPT) and the Big Five personality framework.
We focus on GPT-4o, comparing its behaviour to human baselines and earlier models.
Our findings reveal that GPT-4o exhibits higher Conscientiousness and Agreeableness traits compared to human averages,
while functioning as a risk-neutral rational agent in prospect selection.
Interventions on GPT-4o's Big Five traits, particularly Openness, significantly influence its risk propensity,
mirroring patterns observed in human studies.
Notably, Openness emerges as the most influential factor in GPT-4o's risk propensity, aligning with human findings.
In contrast, legacy models like GPT-4-Turbo demonstrate inconsistent generalization of the personality-risk relationship.
This research advances our understanding of LLM behaviour under risk and elucidates the potential and limitations of personality-based interventions in shaping LLM decision-making.
Our findings have implications for the development of more robust and predictable AI systems such as financial modelling.
\end{abstract}

\section{Introduction}
\label{sec:introduction}
Large language models (LLMs) have emerged as powerful autonomous agents demonstrating versatility across diverse domains \citep{wang2024survey}
such as software engineering \citep{cognition2024introducing, xia2024agentless, qian2024chatdev, qian2023communicative},
financial modelling \citep{lakkaraju2023llms, ding2024large, yu2024finmem},
and multi-agent simulations of emergent human-like behaviour \citep{park2023generative, park2024generative, vallinder2024cultural, park2024generative}.
In the financial sector, agent-based modelling is already utilized in banking \citep{hamill2025agent},
and with the advancement of LLMs, we anticipate an increased application in financial simulations \citep{li2024econagent,zhang2024ai,gao2024simulating}.
The success of LLMs can be attributed to their ability to efficiently leverage knowledge from vast text corpora and their proficiency in natural language understanding and generation~\citep{bubeck2023sparks}.
These capabilities have led to the development of sophisticated agents with advanced decision-making abilities, incorporating linguistic-based modules for planning, memory, perception, and action~\citep{xi2023rise,park2023generative}.

Despite their growing utility and complexity, a comprehensive understanding of LLMs' decision-making processes,
particularly in the context of risk, remains an emerging area of research~\citep{binz2023using, hagendorff2023machine, ross2024llm, binz2024centaur}.
Concurrently, evidence suggests that LLMs can be prompted to exhibit specific personality traits,
such as those defined by the OCEAN model (Openness, Conscientiousness, Extraversion, Agreeableness, Neuroticism)~\citep{serapio2023personality}
and can be prompted to role-play with varying personalities~\citep{serapio2023personality, jiang2024evaluating}.
This finding raises intriguing questions about the interplay between personality traits and decision-making processes in LLMs,
especially in domains where risk-tolerance and personality characteristics play crucial roles, such as financial simulations.

While extensive research has been conducted on the relationship between human personality
and risk behaviour~\citep{nicholson2005personality, soane2005risk, weller2012honest, boyce2016individual, rustichini2016toward, oehler2018relationship, de2019personality, joseph2021personality, highhouse2022risk},
the analogous relationship in LLMs remains unexplored.

Our study aims to address this knowledge gap by investigating the relationship between personality traits and risky decision-making in LLMs.
Specifically, we pose the following research question:
\textit{Do LLMs generalize the relationship between personality traits and risk-based decision-making in a manner analogous to humans?}
To explore this question, we employ two well-established frameworks:
the cumulative prospect theory (CPT) from behavioural economics \citep{tversky1992advances}
and the Big Five personality factors from psychological testing \citep{goldberg1999broad}.

Our primary contributions are summarised as follows:
\begin{enumerate}
    \item We introduce a novel method to estimating the certainty equivalents of LLMs.
    \item We establish that GPT-4o is a risk-neutral rational agent when evaluating risky prospects.
    Our approach significantly outperforms existing methods \citep{ross2024llm} in terms of stability, particularly for evaluations involving potential losses, thereby enhancing reliability in risk assessments.
    \item Our analysis reveals that GPT-4o exhibits notably higher levels of Conscientiousness and Agreeableness, along with lower levels of Neuroticism, compared to a sample of human personality measurements.
    This insight highlights the nuanced personality profile of advanced LLMs and its implications for user interaction.
    \item Our research illustrates that targeted interventions on the Big Five personality traits of GPT-4o lead to irrational risk-averse and risk-seeking behaviors in risk assessments.
    Notably, interventions aimed at the Openness trait produce risk-propensity patterns analogous to those observed in human subjects,
    indicating than GPT-4o generalises the connection between personality and decision-making processes analogously to humans.
    \item We show that GPT-4-Turbo does generalise the personality-risk relationship.
    The legacy model yields a monotonic personality-risk relationship for personality markers with variable qualifiers but not across antonymic personality markers.
    \item We establish that Openness is the most significant personality trait for predicting risk-propensity in LLMs, aligning with findings from human behavioural studies \citep{rustichini2016toward}.
\end{enumerate}

Our approach consists of three main steps.

\begin{enumerate}
\item We assess the personality of GPT-4o by prompting it to self-report answers to the IPIP-NEO-300 Personality Inventory~\citep{goldberg1999broad}.
This inventory decomposes personality into the Big Five personality traits.
\item We intervene on GPT-4o's Big Five personality traits over several levels of trait intensity using a prompting methodology. We refer to these LLMs as \textit{personified}.
\item We investigate risk based decision making in GPT-4o and explore the relationship between induced personality traits and risk-based decision making.
Specifically, we measure GPT-4o's risk-propensity by estimating its CPT model parameters based on the model's preferences for certain outcomes versus risky gambles.
Next, we employ linear models to estimate the significance of the relationship between the induced personality traits on both risk-aversion and loss aversion CPT parameters.
\end{enumerate}

The structure of our paper is as follows: Section \ref{sec:related-work} provides an overview of key literature relevant to estimating CPT parameters in
LLMs, reviewing works from machine learning, psychology, and economics.
In Section \ref{sec:methodology}, we detail our approach to estimating the certainty equivalents of personified LLMs.
Section \ref{sec:results} presents our findings, which include estimates of the Big Five personality trait for GPT-4o,
CPT parameter estimates for GPT-4o, and CPT parameter estimates for personified versions of GPT-4o and GPT-4-Turbo.
Finally, Section \ref{sec:discussion-and-conclusion} offers our concluding remarks and discusses the implications of our research.

\section{Related works}
\label{sec:related-work}
We review related works across machine learning, psychology, and economics in three key areas:
psychometric assessment of personality in humans and LLMs,
the analysis of risk-propensity through the lens of prospect theory (PT),
and the implementation of in-context learning interventions aimed at modulating the personality of LLMs.

\subsection{Personality and risk-propensity in humans}
\label{subsec:personality-and-risk}

Researchers have demonstrated that human personalities can be explained by several independent factors.
A widely accepted model is the Big Five, which categorizes personalities into five traits, often referred to by the acronym OCEAN (Openness, Conscientiousness, Extraversion, Agreeableness, and Neuroticism) \citep{mccrae1987validation, digman1989five, mccrae1992introduction, costa1992normal, saucier1996language}.
An individual's alignment with these factors is measured through their responses to questions associated with specific personality facets for each trait \citep{costa1992normal, goldberg1999broad}.
For instance, the IPIP-NEO inventory \citep{goldberg1999broad}, extensively used in this work, contains several facets for each trait.
Taking Conscientiousness as an example, its first four facets are: \textit{Self-efficacy}, \textit{Orderliness}, \textit{Dutifulness}, and \textit{Achievement-Striving}
(an exhaustive list of personality traits and their associated personality facets is given in Table~\ref{tab:big_five_traits}.).

The following section examines how the Big Five personality traits correlate with risk-propensity in human subjects:

\textbf{Openness to Experience}: The personality trait of Openness is a consistent predictor of risk-taking behaviors in humans.
Research studies have shown its significant correlation with increased risk-propensity across various scenarios.
For instance, \citet{nicholson2005personality} identify Openness as a strong predictor of higher risk-propensity.
\citet{rustichini2016toward} further delineate that Openness is positively associated with risk-taking in contexts where gains are involved, but negatively associated in scenarios involving potential losses.
A meta-analysis conducted by \citet{highhouse2022risk} substantiates these findings, reporting that Openness accounts for 22\% of the variance in risk-propensity.
This indicates a substantial impact of Openness on individual's risk-taking behaviors.
Focusing on investment decisions, \citet{de2019personality} reveal that individuals scoring high on Openness tend to adopt riskier investment strategies.

\textbf{Extraversion}: Multiple studies report a positive correlation between extraversion and risk-propensity.
\citet{nicholson2005personality} and \citet{oehler2018relationship} find that individuals high in extraversion tend to engage in more risk-taking behaviours.

\textbf{Neuroticism}: Research indicates an inverse relationship between neuroticism and risk-taking.
\citet{nicholson2005personality} and \citet{soane2005risk} find that higher levels of neuroticism are linked to lower risk-taking.
Further, \citet{oehler2018relationship} add that neurotic individuals exhibit greater risk aversion, particularly in investment contexts.

\textbf{Agreeableness}: Higher agreeableness generally correlates with lower risk-taking tendencies.
Studies by \citet{nicholson2005personality} and \citet{joseph2021personality} indicate that agreeable individuals are less inclined toward risk.
Additionally, \citet{soane2005risk} observe that agreeableness is inversely related to risk-taking.

\textbf{Conscientiousness}: Conscientiousness shows a complex relationship with risk.
While \citet{nicholson2005personality} suggest lower conscientiousness may correspond with higher risk-taking,
\citet{weller2012honest} specifically associate low conscientiousness with greater risk-taking to achieve gains.
\citet{boyce2016individual} highlight conscientious individuals’ strong reactions to income losses, signifying pronounced loss aversion.
In investment contexts, \citet{oehler2018relationship} note that high conscientiousness aligns with greater risk aversion.

\subsection{Machine psychology}\label{subsec:machine-psychology}

Emergent cognitive behaviours in LLMs cannot be studied under the typical train-and-test machine learning paradigm since these behaviours are not explicitly coded for during training~\citep{hagendorff2023machine}.
Instead, tests from behavioural psychology can be used to reveal characteristic behaviour of black-box LLMs without necessarily understanding how the behaviour was learnt.
Authors caution that care should be taken not to generalise the results of self-reported psychometric test results beyond any particular system prompt \citep{rottger2024political}.
Nevertheless, several works have shown that induced personalities generalise to new tasks~\citep{serapio2023personality, ross2024llm}.

\citet{binz2023using} reported the existence of human-like cognitive biases in GPT-3.
However, more recent studies by \citet{hagendorff2023human} and \citet{chen2023emergence} noted that these cognitive biases have disappeared in the latest generation of LLMs (post GPT-3.5).
These models act as rational agents even without the addition of chain-of-thought prompting.
They have also shown that LLMs answer questions using short-cut learning~\citep{geirhos2020shortcut} or memorisation~\citep{carlini2019secret, carlini2021extracting, hartley2023neural}.
It is hypothesised that popular psychometric tests are in the training data, and due to the uniqueness of their data,
they are likely memorised~\citep{carlini2019secret, carlini2021extracting}.
These behaviours can be mitigated by using procedurally generated prompts.

\citet{salewski2024context} demonstrated that LLMs improved task performance when given personas.
In a multi-armed bandit task, they show that, LLMs impersonating children of various ages mimicked human-like developmental exploration.
In a reasoning task, LLMs portraying domain experts outperformed those assuming non-expert roles.

\subsubsection{Prospect theory and LLMs}

\citet{kahneman1979prospect} introduced prospect theory, a description of how humans make decisions under risk.
PT challenges the assumptions of expected utility theory (EUT)~\citep{von2007theory} by focusing on observed human behaviour rather than prescriptive models of rational choice.
In \citet{kahneman1979prospect} they observed several irrational behaviours which deviated from EUT.
Notably, they found that humans exhibit a distorted perception of probability, overweighting low probabilities and underweighting high probabilities.
Furthermore, they demonstrated that individuals tend to be risk-averse when facing potential gains but become risk-seeking when confronted with potential losses.
Perhaps most significantly, their research revealed that humans are more sensitive to losses than to equivalent gains, a phenomenon known as loss aversion.
cumulative prospect theory~\citep{tversky1992advances} introduced a power-law model to quantify the irrational decision making behaviour observed in \citet{kahneman1979prospect}.
(The model is explained in detail in Section \ref{sec:methodology}).

More recently, several researchers have investigated decision-making in LLMs with respect to prospect theory.
\citet{binz2023using} submitted GPT-3 to a number of vignettes and tasks from the cognitive psychology literature.
They found that GPT-3 showed several human-like cognitive biases from PT: certainty effect \footnote{The certainty effect is the tendency of agents to prefer certain outcomes to risky outcomes.},
overweighting effect\footnote{The overweighting effect is the tendency of agents to overweight the probability of rare outcomes.},
and the framing effect\footnote{The framing effect describes the influence of the presentation of outcomes on decision-making.}.
In subsequent work, \citet{binz2023turning} improved the alignment between LLama 2--65B~\citep{touvron2023llama} and human decision-making by training a
logistic regression model on LLama 2 embeddings of risky prospects from the choices13k dataset~\citep{peterson2021using}.
They also achieved increased performance on a hold-out task.
However, the limitation of this work is that the model is no longer generative.
Concurrently~\cite{ross2024llm}, measured the CPT parameters of GPT-4o and GPT-4-Turbo and found that cognitive biases in these
more recent models are reduced.
The authors also show that risk-propensity is sensitive to qualitative personas.
However, a quantitative comparison with risk personas in humans cannot be made since these profiles are not based on established personality models.
In our work we investigate the relationship between established personality models and risk-propensity in LLMs,
and examining the similarities between these relationships in humans.

\subsection{Personality prompting}\label{subsec:personality-prompting}

Recent research has extensively explored the induction and measurement of personalities in LLMs. For example, \citet{jiang2024evaluating} employed self-reported psychometric testing to assess LLMs' personalities along the Big Five traits.
They developed a method to induce these traits in LLMs and demonstrated that the induced personalities generalised to vignette experiments.
Similarly, \citet{serapio2023personality} reported on LLM personality types using psychometric tests.
Their method allowed for independent induction of personalities along each trait, with controllable levels for the intensity of each personality trait.
The authors validate their method by measuring the correlation between induced personality trait levels and the LLMs' responses to the IPIP-NEO-300 inventory for each trait.
Furthermore, they observed that these personalities generalised well to downstream tasks such as generating social media posts.

Several studies have explored personality characteristics beyond the Big Five personality traits.
\citet{li2022does} reported that LLMs generally score higher than the human average for toxic personality traits,
specifically the Dark Triad~\citep{furnham2013dark}.
The authors successfully implemented interventions on these traits using in-context learning.
\citet{miotto2022gpt} assessed GPT-3's personality using the 60-item Hexaco questionnaire~\citep{ashton2009hexaco},
revealing multiple personalities at different sampling temperatures.
\citet{coda2023inducing} found that GPT-3.5 displays higher than human average anxiety on the State-Trait Inventory for Cognitive and Somatic Anxiety (STICSA) Questionnaire~\citep{ree2008distinguishing}.

The validity of self-reported personality testing in LLMs has recently been questioned.
\citet{suhr2023challenging} demonstrated that GPT-3.5 and LLama 2 do not show consistency in their evaluation of personality inventories.
For example, their factor analysis revealed that responses to the Big-Five-Inventory 2 (BFI-2) do not show a simple structure for the first five factors of variation.
However, they found that the component loadings for GPT-4-Turbo are separable for each factor, indicating consistent personality traits across questioning for this more advanced model.

\section{Methodology}\label{sec:methodology}
\subsection{Estimating CPT parameters}\label{subsec:estimating-risk-propensity}

Prospect theory models decision-making under risk,
describing how agents evaluate and choose between risky prospects.
In this framework, agents select prospects offering the highest utility.
The utility of a prospect is determined by a function that combines subjective outcome probabilities with their subjective values, as follows:
\begin{equation}
  u(P) = \sum_{i=1}^{n}w(p_i)v(x_i)
    \label{eq:utility}
  \end{equation}
where $x_i$ represents an outcome with an associated probability $p_i$.
The set $P$, defined as $\{x_i, p_i\}_{i=1}^n$ encompasses all the possible outcomes and their corresponding probabilities within a prospect.
The function $w(p_i)$ serves as the probability weighting function, transforming objective probabilities into decision weights.
Complimenting this, $v(x_i)$ acts as the value function, assigning a subjective value to each outcome.

The value function and the weighting function are given by equation \eqref{eq:equation-value} and equation \eqref{eq:weight} respectively,
and are parameterised by the CPT parameters $\vec{\theta}=(\alpha, \beta, \lambda, \gamma)$.
\begin{equation}
v(x) =
\begin{cases}
x^\alpha & \text{if } x \geq 0 \\
-\lambda (-x)^\beta & \text{if } x < 0
\end{cases}\label{eq:equation-value}
\end{equation}
\begin{equation}
w(p) = \frac{p^\gamma}{(p^\gamma + (1 - p)^\gamma)^{1/\gamma}}
  \label{eq:weight}
\end{equation}

The CPT parameters describe the risk-propensity of an agent as follows:
\begin{itemize}
  \item Gain sensitivity ($\alpha$) - Diminishing sensitivity to gains (risk-averse) for $0<\alpha<1$.
  \item Loss sensitivity ($\beta$) - Diminishing sensitivity to losses (risk-seeking) for $0<\beta<1$.
  \item Loss aversion ($\lambda$) - Losses loom larger than equivalent gains for $\lambda>1$.
  \item Probability sensitivity ($\gamma$) - Individuals overweight small probabilities and underweight large probabilities for $0 < \gamma < 1$ .
  For gains and losses the probability sensitivity is denoted by $\phi_+$, $\phi_-$ respectively.
\end{itemize}
We estimate an agent's CPT parameters by analysing its preferences for certain outcomes to risky prospects.
For instance, consider a choice between a risky prospect
\textit{10\% chance of \$0, 90\% chance of \$100} and a certain outcome.
A risk-averse agent prefers a certain outcome below the risky prospect's expected value,
while a risk-seeking agent favours one about it.
The certain outcome at which the agent is indifferent is called the certainty equivalent.

Our approach to determining CPT parameters is as follows.
First, we ask the agent to return its certainty equivalent to each prospect in the dataset (see Section \ref{section:estimation} for dataset details).
Second, we perform non-linear regression on these observed certainty equivalents.
We solve for the CPT parameters that minimize the regression problem in equation \eqref{eq:regression_1}.

The optimization procedure is initialized using median human CPT parameters from \citet{tversky1992advances}:
$\vec{\theta}_0=(\alpha=0.88, \beta=0.88, \lambda=2.25, \phi+=0.61, \phi_-=0.69)$.
To ensure $u$ accurately represents human behaviour, we constrain all parameters to non-negative values.
We employ Nelder-Mead optimization for the regression \citep{nelder1965simplex}.
\begin{equation} \vec{\theta}^* = \min_{\vec{\theta}} \frac{1}{n} \sum_{i=1}^n \left( c_i - v^{-1}[u(P_i | \vec{\theta})] \right)^2 \label{eq:regression_1}\end{equation}
\begin{equation*} \text{subject to} \quad \theta_j > 0, \quad \forall j \in \{1, \ldots, m\} \end{equation*}
where $\vec{\theta}$ are the agent's CPT parameters,
$c_i$ is the agent's certainty equivalent to prospect $P_i$, $u(P_i | \vec{\theta})$ is the utility of a prospect given CPT parameters
$\vec{\theta}$, $v^{-1}$ is the inverse function of $v$,
$n$ is the number of risky prospects presented to the agent,
and $m$ is the number of CPT model parameters.
\subsection{Estimating the CPT parameters of LLMs}\label{section:estimation}
\citet{ross2024llm} employed a method for estimating an LLM's certainty equivalents to risky prospects, adapted from \citet{tversky1992advances}.
The procedure begins by presenting the LLM with a prospect, along with seven certain outcomes.
These outcomes are logarithmically spaced between the minimum and maximum values of the prospect's possible outcomes.
Log-spaced outcomes permit estimates of the CPT values across a multiple orders of magnitude.
In the first round, the LLM indicates which certainty equivalents it would accept or reject.
The second round refines the estimate using linearly spaced certainty equivalents between the accepted/rejected outcomes from the first round.
The final certainty equivalent is estimated as the midpoint of the accepted/rejected outcome.
The process is repeated for all prospects, and the CPT parameters estimated using a non-linear regression.

We identified three primary issues with the previous approach.
First, GPT-4o frequently misinterprets the magnitudes of certainty equivalents of prospects with negative expected values.
Second, the log-spacing of certainty equivalents led to less precise estimates when dealing with large outcomes.
This occurred because the spacing between larger outcomes was disproportionately wide compared to the spacing between smaller outcomes.
Third, interventions on personality traits introduced instabilities.
When we attempted to intervene on personality traits, the model often exhibited unstable behaviour,
resulting in uniform acceptance or rejection of all certainty equivalents.

Our methodology leverages direct reporting of certainty equivalents by the LLM
(see the system prompt in Figure~\ref{fig:system-prompt} and the user prompt in Figure~\ref{fig:user-prompt})\footnote{All experiments are ran with temperature set to $1.0$.}.
To differentiate between prospects with positive and negative expected values,
we specifically request the \textit{least positive} or \textit{most negative} cash amounts.
We infer the certainty equivalents for all prospects over 15 runs in each experiment.
We ensure consistency in token sampling for each prospect by using a fixed random seed in each run.
After measuring the agent's certainty equivalents, we estimate the agent's CPT parameters using equation~\eqref{eq:regression_1} .
We compare the values of certainty equivalents using our approach and \citet{ross2024llm} in Figure~\ref{fig:ce-ev-our-work} and Figure~\ref{fig:ce-ev-previous} in the appendix respectively.
We find that our approach leads to a reduction in the number of outliers for prospects with negative expected values.
We believe that the simplicity of our approach reduces the instabilities present in their approach.

We use prospects from two datasets, $D_A$ and $D_B$, to estimate the CPT parameters of agents.
Dataset $D_A$ consists of all 56 prospects from \citet{tversky1992advances}.
Conversely, dataset $D_B$ comprises an equal number prospects containing gains and losses (mixed) which are randomly sampled from choices13k \citep{peterson2021using}.
The prospects in $D_A$ and $D_B$ are given in Table~\ref{tab:da-prospects} and Table~\ref{tab:db-prospects} in the appendix respectively.
Figure~\ref{fig:dataset-probability-distribution} shows the distribution of probabilities for the second outcome in prospects in $D_A$ and $D_B$.
Figure~\ref{fig:dataset-expected-value-distribution} shows the distribution of the expected values of the prospects in $D_A$, $D_B$.

Dataset $D_A$ contains the median certainty equivalents for 25 graduate students.
Individual certainty equivalents are not available.
We compare the agent's certainty equivalents with the median human certainty equivalents directly.
However we use $D_B$ to estimate the agent's CPT parameters.
This is because $D_B$ contains mixed prospects which are required to estimate the loss aversion parameter $\lambda$.
Whilst mixed prospects are not available in \citet{tversky1992advances}.
We hypothesize that dataset $D_A$ was used in \citet{ross2024llm} to estimate the CPT parameters.
This inference is based on the observation that their loss aversion parameter estimates are significantly less precise compared to the estimates for other CPT parameters,
suggesting that the optimization process converged independently of the loss aversion parameter.
Dataset $D_B$ offers an additional advantage: it mitigates the risk of prediction biases stemming from LLMs' potential memorisation of in-sample training data \citep{carlini2019secret,hartley2023neural}.
This benefit arises because the choices13k dataset does not include human-estimated certainty equivalents,
thus reducing the likelihood of the model simply recalling pre-existing human judgments.

\begin{figure}[!htbp]
  \centering
  \begin{tcolorbox}[colback=gray!10, colframe=black, width=.9\textwidth, title=Prospect System prompt]
      \ttfamily
You will be given a gamble. \\
\\
What is the \{\{least positive / most negative\}\} amount of money that you would accept instead of taking the gamble? \\
\\
Give your answer in the following format \\
\\
reason: reason for your choice \\
answer: amount
  \end{tcolorbox}
  \caption{System prompt to estimate the certainty equivalent to a risky prospect. We define the certainty equivalent by asking
  for the \textit{least positive} or \textit{most negative} amount of money accepted.}
  \label{fig:system-prompt}
\end{figure}

\begin{figure}[!htbp]
  \centering
  \begin{tcolorbox}[colback=gray!10, colframe=black, width=.9\textwidth, breakable, title=Prospect user prompt]
      \ttfamily
\{\{Outcome 1\}\} dollars with \{\{100 - p\}\}\% probability and \{\{Outcome 2\}\} dollars with \{\{p\}\}\% probability. \\
  Lets think about this step by step
  \end{tcolorbox}
  \caption{User prompt for a prospect for an LLM primed by the system prompt in Figure~\ref{fig:system-prompt}, where \textrm{Outcome 1} and \textrm{Outcome 2} are the lowest and highest outcomes in the prospect and \textrm{p} is the probability as a percentage of \textrm{Outcome 2} occuring.}
  \label{fig:user-prompt}
\end{figure}
\subsection{Measuring the effect of personality interventions on the risk-propensity of LLMs}\label{subsec:personality-interventions-on-the-risk-propensity-of-llms}

This section outlines our methodology for modifying personality traits LLM and assessing the resulting changes in their risk-propensity.
Our technique involves instructing LLMs to answer questions about risk prospects in alignment with predefined personality templates from \cite{serapio2023personality}.
These templates are designed to represent specific personality traits at varying levels of intensity,
allowing us to systematically alter the LLM's behavioural responses.

Recent studies have demonstrated that the \textit{personalities} of LLMs can be measured through psychometric testing~\citep{karra2022estimating}.
More recently, researchers have successfully induced personality traits in LLMs using in-context learning techniques~\citep{serapio2023personality, jiang2024evaluating}.
Notably, these induced personalities have been shown to generalize beyond psychometric testing to other tasks~\citep{serapio2023personality, jiang2024evaluating}.
For instance, induced personality traits in Flan-PaLM 540B exhibited moderate-to-strong correlations with personality levels observed in generated social media posts~\citep{serapio2023personality}.

We begin by making a comparison of GPT-4o's personality traits with a human sample \citep{johnson2005ascertaining}.
First we measure GPT-4os responses to the IPIP-NEO-300 personality inventory~\citep{goldberg1999broad}.
This inventory comprises 300 questions designed to assess an individual's association with personality facets on a 1--5 scale.
Scores for the Big Five personality traits are calculated as the sum of responses for facets belonging to each trait.
For example, the question \textit{Worry about things} correlates with the facet \textit{Anxiety},
which is associated with the personality trait \textit{Neuroticism} (see Table \ref{tab:big_five_traits} for a comprehensive list of facets and personality traits in the IPIP-NEO-300).
Then, we compare the measured traits against a human sample of UK citizens aged 30 and above~\citep{johnson2005ascertaining}.

Next, we investigate the relationship between personality and risk-propensity by intervening
on the Big Five personality traits and measuring the resulting CPT parameters.
To accomplish this, we modify the system prompt by prepending it with an instruction to adopt
a persona based on a single trait, following the approach outlined by \citet{serapio2023personality}.

The structure of our modified system prompt is depicted in Figure \ref{fig:personality_system_prompt}.
Our intervention incorporates two primary elements: bipolar adjective markers and a Likert item.
Research by \citet{goldberg1999broad} demonstrated that bipolar adjective markers (such as intelligent-unintelligent)
effectively characterize the Big Five personality traits.
Building on this, \citet{serapio2023personality} employed these bipolar-adjective markers to manipulate personality
traits in LLMs.
In this approach, the Likert item regulates the intensity of the marker (e.g., \textit{extremely}) on a 1--9 scale
(We use the Likert scale shown in Figure \ref{fig:qualifiers}).
Each marker represents a specific facet of a personality trait.
The markers are categorised into low-level (levels 1--3) and high-level (levels 5--9) facets,
with level 4 serving as a neutral point that combines both low and high-level markers.
A comprehensive list of these markers is provided in Table \ref{tab:low-high-markers}.
This methodology enables us to intervene on each personality trait and its corresponding level independently.

\begin{figure}
    \centering
    \begin{tcolorbox}[
        colback=gray!10,
        colframe=black,
        width=0.9\textwidth,
        title=Personality intervention system prompt
    ]
        \ttfamily
        For the following task, respond in a way that matches this description: \\
        "I'm very unsure, very messy, very irresponsible, very lazy, very undisciplined, very impractical, very extravagant, very disorganized, very negligent, very careless."
    \end{tcolorbox}
   \caption{An example of system prompt to intervene on the Big Five personality trait of Conscientiousness.}
   \label{fig:personality_system_prompt}
\end{figure}

In our final analysis, we compare how the Openness personality trait relates to risk-propensity in both GPT-4o and GPT-4-Turbo.
We chose to focus on Openness because previous research has identified it as the most significant personality trait in
explaining human risk-propensity \citep{highhouse2022risk}.
Our methodology involves fitting linear models to the CPT parameters,
using the level of Openness as a independent variable.
We then examine the direction (sign) of the resulting model coefficients and assess their statistical significance.
This approach enables us to quantify and contrast the influence of Openness on risk-taking behaviour across these two language model variants,
while also comparing our findings to established human behavioural patterns reported in the literature.

\section{Results}\label{sec:results}

\subsection{GPT-4o is a risk-neutral rational agent}\label{subsec:gpt-4o-is-a-rational-agent}

We present a comprehensive analysis of decision-making behaviour in GPT-4o,
comparing it with risk-neutral rational agents and human subjects across a range of prospects.

\begin{figure}[!htbp]
  \centering
  \includegraphics{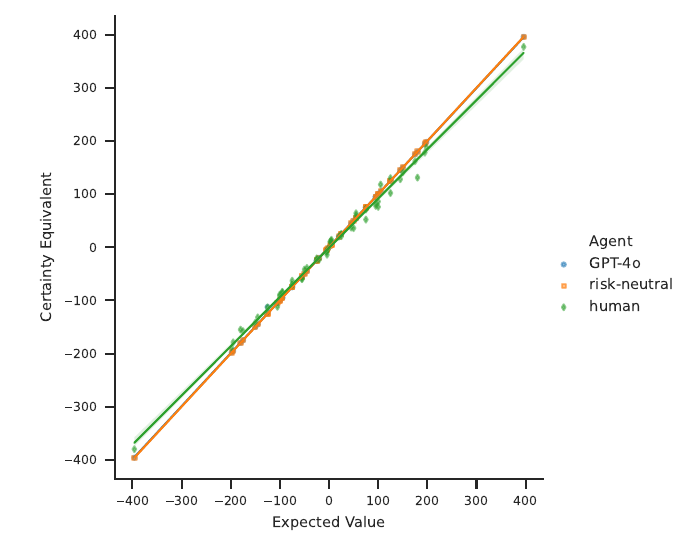}
  \caption{A comparison of certainty equivalents and expected values for non-mixed prospects from the dataset $D_A$ across different agents.
  Linear regression lines for each agent are displayed, with shaded regions representing 95\% confidence intervals.
  Note that that the results for GPT-4o and risk-neutral agents are near indistinguishable.}
  \label{fig:ce-eq-agents-comparison}
\end{figure}

Figure~\ref{fig:ce-eq-agents-comparison} shows the certainty equivalents for
GPT-4o, risk-neutral rational and human agents over a range of non-mixed prospects ($D_A$).
We show the median certainty equivalents for each agent in Table~\ref{tab:certainty_equivalents}.
We use the methodology detailed in Section~\ref{subsec:estimating-risk-propensity}
Notably, GPT-4o exhibits certainty equivalents that are either identical or closely aligned
with those of rational risk neural agents for nearly all prospects (i.e.\ certainty equivalents are equal to expected values of the prospects).
This similarity suggests that GPT-4o's decision-making process closely adheres to the principles
of expected utility theory, where the utility function is directly proportional to the outcome.

To further investigate the decision-making characteristics of GPT-4o,
we estimated its CPT parameters using a set of mixed prospects ($D_B$).
Figure~\ref{fig:vanilla-cpt-parameters} presents these estimates,
over 10 random seeds.

Our analysis of the CPT parameters reveals several key findings:

\begin{itemize}

\item The introduction of mixed prospects induces a minimal level of risk aversion for gains in GPT-4o, signified by $\alpha < 1$.
The results contrasts with the results observed in non-mixed prospects (Table~\ref{tab:certainty_equivalents} in the appendix).

\item The estimates for the CPT parameters $\beta$, $\gamma$, $\lambda$, $\phi_+$ and $\phi_-$ are statistically indistinguishable from unity
within the bootstrapped 95\% confidence intervals.
This result strongly indicates that GPT-4o's decision-making behaviour aligns closely with expected utility theory.

\item Contrary to previous findings~\citep{ross2024llm} for GPT-4 and GPT-4-Turbo,
which demonstrated underweighting of low probabilities and overweighting of high probabilities,
our results show that GPT-4o's probability weighting parameters ($\phi_-$ and $\phi_+$)
are equal to 1.
This confirms an absence of probability distortion in GPT-4o's decision-making process.

\end{itemize}

These findings extend the recent work by \citet{ross2024llm},
which demonstrated an increasing trend towards rationality in decision-making for GPT-4 and GPT-4-Turbo.
Our results provide further evidence that more recent LLMs
exhibit enhanced rational decision-making capabilities.

\subsection{GPT-4o exhibits human-like risk-personality patterns}\label{subsec:gpt-4o-shows-human-like-risk-personality-patterns}

In this section, we investigate the relationship between personality traits and risk propensity in GPT-4o,
comparing it with human behaviour.
Our findings reveal both similarities and differences,
contributing to the understanding of LLMs' decision-making under risk.

\begin{figure}[h]
  \centering
  \begin{subfigure}[t]{0.45\textwidth}
    \centering
    \includegraphics[width=\textwidth]{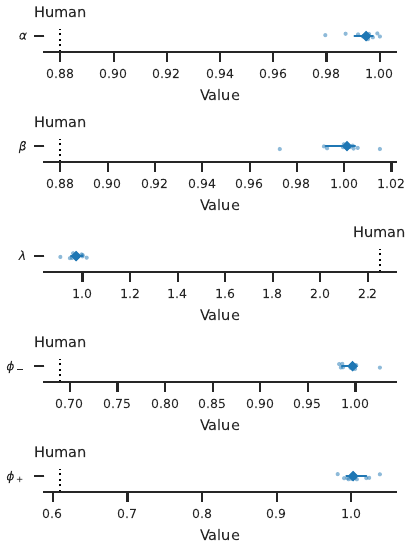}
    \subcaption{CPT parameter estimates for humans and GPT-4o (15 runs).
    We compare the GPT-4o estimates with the median estimates for 25 human participants~\citep{tversky1992advances}.}
    \label{fig:vanilla-cpt-parameters}
  \end{subfigure}
  \hfill
  \begin{subfigure}[t]{0.45\textwidth}
      \centering
      \includegraphics[width=\textwidth]{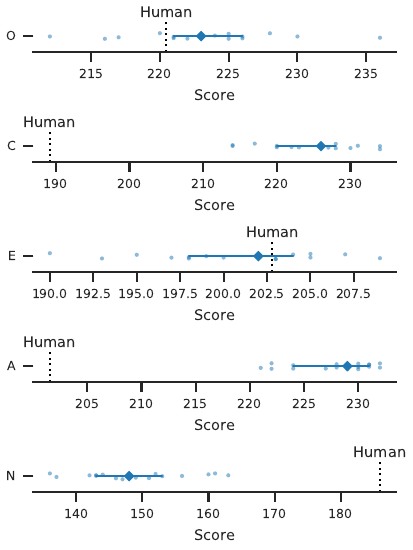}
      \subcaption{Big Five personality trait scores for GPT-4o (25 runs) and humans in response to the IPIP-NEO-300 Inventory.
      GPT-4o scores are compared with the mean score per trait from a human sample \citep{johnson2005ascertaining}.}
      \label{fig:personality-estimates}
  \end{subfigure}
  \caption{A comparison of CPT parameters and personality trait score estimates for GPT-4o and humans.
  A unique random seed is used to control random token sampling in each run (denoted by circles).
  Median values (denoted by diamonds) and bootstrapped 95\% confidence intervals are given for GPT-4o results.}
   \label{fig:gpt-4o-personality-risk}
\end{figure}

Figure~\ref{fig:personality-estimates} illustrates GPT-4o's scores for each personality trait on the IPIP-NEO-300 personality inventory.
Prior research has established Openness as the most influential personality trait in human risk-taking behaviour \citep{highhouse2022risk}.
\citet{rustichini2016toward} demonstrated that Openness positively correlates with risk-taking for gains
and negatively correlates with risk-taking for losses in humans.
Our analysis, as shown in Figure~\ref{fig:personality-estimates},
indicates that GPT-4o exhibits higher Openness scores compared to the human sample \citep{johnson2005ascertaining}.
Correspondingly, Figure~\ref{fig:personality-cpt-estimates} reveals that GPT-4o demonstrates increased risk-taking
for gains and decreased risk-taking for losses.
These findings suggest a parallel relationship between Openness and risk-taking behaviour in both GPT-4o and humans,
highlighting a potential similarity in decision-making processes between AI and human cognition.

While Openness plays a primary role,
other personality traits: Conscientiousness, Agreeableness, Extraversion, and Neuroticism—also influence risk propensity, albeit to a lesser extent.
Previous studies by \citet{nicholson2005personality} and \citet{joseph2021personality} indicate that lower Conscientiousness
and higher Agreeableness are associated with higher and lower risk-taking, respectively, in humans.
Our results, as depicted in Figure~\ref{fig:personality-estimates},
show that GPT-4o exhibits significantly higher levels of both Conscientiousness and Agreeableness compared to the human sample.
Considering the low level of risk-aversion given by $\alpha$ in Table~\ref{fig:vanilla-cpt-parameters},
these findings suggest that the relationships between Conscientiousness, Agreeableness, and risk-taking observed in humans are mirrored in GPT-4o,
further supporting the notion of analogous decision-making processes.

However, our study also reveals an intriguing divergence.
Previous research \citep{nicholson2005personality,soane2005risk,oehler2018relationship} has shown that lower
Neuroticism is associated with higher risk-taking in humans.
In contrast, our results indicate that GPT-4o exhibits a significantly lower level of Neuroticism than the human sample (Figure~\ref{fig:personality-estimates}),
yet demonstrates a low level of risk-aversion,
as evidenced by $\alpha$ in Table~\ref{fig:vanilla-cpt-parameters}.
This discrepancy suggests that the relationship between Neuroticism and risk-aversion observed in humans is not mirrored in GPT-4o.

\subsection{Openness in GPT-4o correlates with risk-propensity}\label{subsec:openness-in-gpt-4o-correlates-with-risk-propensity}

In this experiment we intervene independently on each of the Big Five personality traits of GPT-4o and measure the CPT parameters across prospects from the dataset $D_B$.
Figure~\ref{fig:personality-cpt-estimates} shows the variation in CPT parameters by trait level across each personality trait.

Our primary finding reveals a correlation between Openness and risk-taking behaviour for gains ($\rho=0.63$),
and a correlation for risk aversion for losses ($\rho=0.44$,
see Table~\ref{tab:correlations-4o} for correlation coefficients across all personality traits).
This result is in agreement with established risk taking patterns in human subjects~\citep{nicholson2005personality, rustichini2016toward,de2019personality}.
We also find that extraversion is significantly correlated with risk taking for gains.
This behaviour is in agreement with studies on risk aversion in humans \cite{nicholson2005personality, oehler2018relationship}.
We also find that Openness has the highest correlation with the risk aversion parameters.
This observation corroborates the findings of~\cite{rustichini2016toward}, who reported that Openness the primary personality trait relating to risk-propensity.
We not observe any statistically significant correlations between the level of any other personality trait and risky aversion for gains and losses.

Although correlations for Openness in GPT-4o are consistent with those observed in human subjects,
we are unable to suggest whether the bias is correct for these parameter estimates.
In fact our results show that we are unable to produce risk-seeking behaviours for gains or risk-averse behaviours for losses in an absolute sense.

This limitation suggests that personality prompting alone is insufficient to elicit these specific risk behaviours in GPT-4o.
We emphasize the importance of this result for researchers designing personality profiles based on the Big Five traits for agents in multi-agent simulations, such as those proposed by \citet{park2023generative}.
\begin{table}[!ht]
  \centering
  \caption{Pearson's correlation coefficients between the level of a personality trait intervention and
  the model's estimated CPT parameter values over non-mixed prospects the dataset $D_B$ for GPT-4o.
  $\alpha$, $\beta$, $\lambda$ are estimated from certainty equivalents using equation~\eqref{eq:regression_1}.
  The significance of coefficients is determined using t-statistics.
  Results marked with $^{*}$/$^{**}$/$^{***}$ have statistical significance at $\alpha=0.05/0.025/0.001$ level respectively.}
  \label{tab:correlations-4o}
  \vspace{0.1cm}
  \begin{tabular}{l S[table-format=-1.2] S[table-format=-1.2] S[table-format=-1.2]}
    \toprule
    \textbf{Trait} & $\alpha$ & $\beta$ & $\lambda$ \\
    \midrule
    Openness & 0.63$^{***}$ & 0.44$^{**}$ & -0.55$^{***}$ \\
    Conscientiousness & -0.25 & -0.06 & -0.30$^{*}$ \\
    Extraversion & 0.28$^{*}$ & 0.05 & -0.46$^{***}$ \\
    Agreeableness & 0.04 & 0.24 & -0.68$^{***}$ \\
    Neuroticism & 0.01 & -0.05 & 0.37$^{***}$ \\
    \bottomrule
    \end{tabular}
  \end{table}

\subsection{Global and local variations in personality-risk-propensity relationships across LLMs}\label{subsec:global-and-local-variations-in-personality-risk-propensity-relationships-across-large-language-models}

We performed a comparative analysis of personality interventions on GPT-4-Turbo and GPT-4o with a focus on the Openness trait,
identified as a critical factor in risk-taking behaviour~\cite{highhouse2022risk}.
Specifically, we fit linear regression models to the each of the CPT parameters for risk sensitivity to the prospects in the dataset $D_B$,
using the Openness level as the independent variable ($y = \omega x$ where $y$ is CPT parameter, $\omega$ is the model weight and $x$ is the Openness level).
The learnt weight for each model and its statistical significance is presented in Table~\ref{tab:coefficients}.
We find that GPT-4-Turbo contradicts established human risk behaviour pattern,
displaying a negative correlation with risk-taking for gains and a positive correlation for losses.

\begin{table}[!ht]
  \centering
  \caption{Weight parameters for linear models mapping the level of Openness to CPT parameters for GPT-4-Turbo and GPT-4o.
  Results marked with $^*$ indicate a significance level of $\gamma=0.05$.}
  \label{tab:coefficients}
  \vspace{0.1cm}
  \begin{tabular}{l S[table-format=-1.3] S[table-format=-1.3] S[table-format=-1.2]}
    \toprule
    \textbf{Model} & $\alpha$ & $\beta$ & $\lambda$ \\
    \midrule
    GPT-4-Turbo & -0.096$^*$ & -0.092$^*$ & -0.67$^*$ \\
    GPT-4o & 0.039$^*$ & 0.052$^*$ & -0.43$^*$ \\
    \bottomrule
  \end{tabular}
\end{table}

\begin{figure}[!ht]
  \centering
  \includegraphics[width=\textwidth]{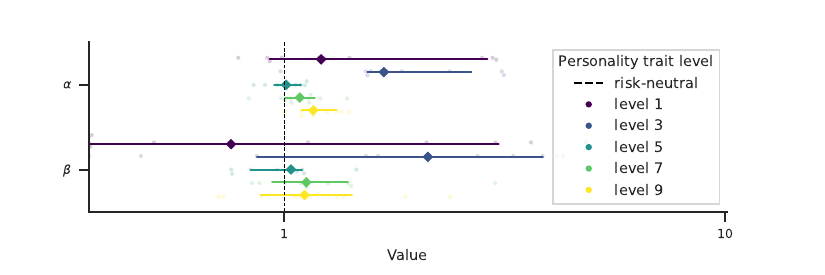}
  \caption{CPT parameter estimates for GPT-4-Turbo with personality interventions on the Openness personality trait are induced at five levels, ranging from low to high.
  CPT parameters are estimated across $D_B$ over 10 runs, each with a unique random seed. Diamond markers represent median values for each parameter,
  with error bars indicating the bootstrapped 95\% confidence interval.}
  \label{fig:gpt-4-alpha-beta}
\end{figure}

To further investigate this discrepancy, Figure~\ref{fig:gpt-4-alpha-beta}
illustrates the variation in $\alpha$, $\beta$ across different levels of Openness for GPT-4-Turbo.
Interestingly, the relationship between personality and risk-taking in GPT-4-Turbo aligns with human behaviour (almost monotonic) for specific subgroups of trait levels: $[1, 3]$ and $[5, 7, 9]$.
We hypothesize that this localized agreement emerges from the independence of personality markers in the two subgroups.
For example, the \textit{Socially Conservative} marker is present only in the first group,
while its antonym marker \textit{Socially Progressive} is present only in the second sub-group (see Table~\ref{tab:low-high-markers} for a complete list of personality trait markers).
risk-propensity is clearly not globally monotonic over these the sub-groups whereas it is for GPT-4o (see Figure~\ref{fig:gpt-4o-personality-risk}).

Our findings imply that GPT-4o exhibits a global mapping of personality to risk-propensity,
whereas GPT-4-Turbo demonstrates a localized mapping that correlates with human data only for specific personality markers.

\section{Conclusion}\label{sec:discussion-and-conclusion}

In this paper, we have presented a comprehensive analysis of risk-propensity in large language models, focusing on GPT-4o.
Our findings reveal that GPT-4o acts as rational risk neutral agent when selecting prospects.
This is evidenced by minimal probability distortion and consistent risk-neutral patterns between certainty equivalents and expected value, marking a significant advancement in LLM capabilities.

A key contribution of our work is the demonstration of a correlation between personality traits, particularly Openness,
and risk-propensity in GPT-4o that mirrors human behavioural patterns.
This finding opens up new avenues for developing risk-propensity calibrated AI agents through personality-based interventions.

Our comparative analysis with GPT-4-Turbo highlights important differences in the global versus local mappings of personality to risk behaviors.
This underscores the need for further research to enhance consistency in AI behavioural modelling across different LLM architectures.

We have also quantified the sensitivity of cpt parameters to prompt-based personality interventions,
providing insights into the malleability of risk behaviour in LLMs.
However, our work also reveals limitations in inducing specific risk behaviors,
pointing to the need for more sophisticated intervention strategies.

Looking ahead, we identify two critical areas for future research.
First, we propose leveraging reinforcement learning techniques to discover optimal prompts that can induce specific risk profiles in LLMs.
Secondly, we advocate for investigating interventions on latent activation directions within open-source LLMs to develop more controllable and interpretable mechanisms for modulating risk-taking behavior,
These advancements will be crucial for creating agentic financial simulations with predictable and adjustable risk profiles.

\section*{Acknowledgements}
We wish to express appreciation to Greig Cowan, Graham Smith, and Zachery Anderson of NatWest Group for the time and support needed to develop this research paper.

\bibliographystyle{icml2014}
\bibliography{library}

\newpage

\appendix
\section{Supplementary results}
\label{sec:a}

\subsection{Median certainty equivalents for non-mixed gambles}\label{subsec:median-certainty-equivalents-for-non-mixed-gambles}

Table~\ref{tab:certainty_equivalents} shows the median certainty equivalents for
GPT-4o, human and rational agents estimated over the non-mixed
prospects in the dataset $D_A$.
The certainty equivalents for GPT-4o and rational
agents identical except for a small minority of values.
This result strongly suggests that GPT-4o behaves as a rational agent
in the non-mixed prospect setting.

\begin{table}[!htb]
  \centering
  \caption{Median certainty equivalents for Rational/Human/GPT-4o agents for non-mixed prospects from \cite{tversky1992advances}.
  Outcome 1 and Outcome 2 are the dollar amounts for the risky prospect, and $p=P(\mathrm{Outcome~2})$.}
  \label{tab:certainty_equivalents}
  \vspace{0.1cm}
  \resizebox{\textwidth}{!}{
  \begin{tabular}{c|c|ccccccccc}
  \toprule
  \textbf{Outcome} 1 & \textbf{Outcome 2} & \multicolumn{9}{c}{\textbf{Median Certainty Equivalents}} \\
  \cmidrule(lr){3-11}
  & & $p=0.01$ & $p=0.05$ & $p=0.1$ & $p=0.25$ & $p=0.5$ & $p=0.75$ & $p=0.9$ & $p=0.95$ & $p=0.99$ \\
  \midrule
  0 & 50 &  &  & 5/9/5 &  & 25/21/25 &  & 45/37/45 &  &  \\
  0 & -50 &  &  & -5/-8/-5 &  & -25/-21/-25 &  & -45/-39/-45 &  &  \\
  0 & 100 &  & 5/14/5 &  & 25/25/25 & 50/36/50 & 75/52/75 &  & 95/78/95 &  \\
  0 & -100 &  & -5/-8/-5 &  & -25/-24/-25 & -50/-42/-50 & -75/-63/-75 &  & -95/-84/-95 &  \\
  0 & 200 & 2/10/2 &  & 20/20/20 &  & 100/76/100 &  & 180/131/180 &  & 198/188/198 \\
  0 & -200 & -2/-3/-2 &  & -20/-23/-20 &  & -100/-89/-100 &  & -180/-155/-180 &  & -198/-190/-198 \\
  0 & 400 & 4/12/4 &  &  &  &  &  &  &  & 396/377/396 \\
  0 & -400 & -4/-14/-4 &  &  &  &  &  &  &  & -396/-380/-396 \\
  50 & 100 &  &  & 55/59/55 &  & 75/71/75 &  & 95/83/95 &  &  \\
  -50 & -100 &  &  & -55/-59/-55 &  & -75/-71/-75 &  & -95/-85/-95 &  &  \\
  50 & 150 &  & 55/64/54 &  & 75/72/75 & 100/86/100 & 125/102/125 &  & 145/128/145 &  \\
  -50 & -150 &  & -55/-60/-55 &  & -75/-71/-75 & -100/-92/-100 & -125/-113/-112 &  & -145/-132/-145 &  \\
  100 & 200 &  & 105/118/105 &  & 125/130/125 & 150/141/150 & 175/162/175 &  & 195/178/195 &  \\
  -100 & -200 &  & -105/-112/-105 &  & -125/-121/-125 & -150/-142/-150 & -175/-158/-175 &  & -195/-179/-195 &  \\
  \bottomrule
  \end{tabular}}
\end{table}

\subsection{Interventions on personality traits}
\label{subsec:inducing-personality-traits}

\begin{figure}[!htb]
    \centering
    \includegraphics[width=\textwidth]{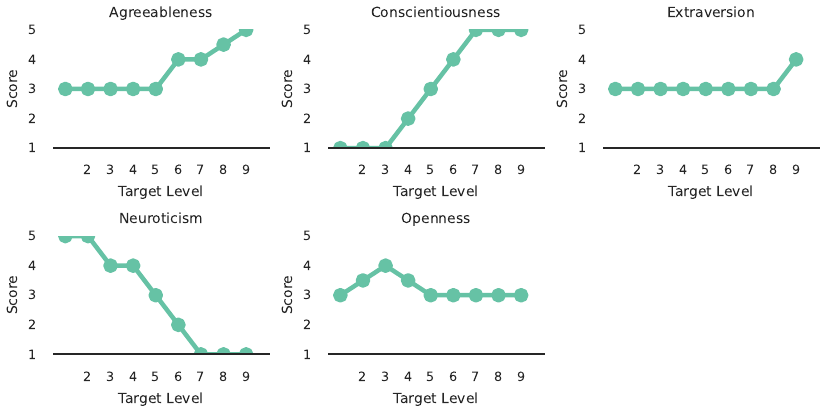}
    \caption{Median scores on the IPIP-NEO-300 personality inventory for GPT-4o with interventions on conscientiousness.}
    \label{fig:induction-level}
\end{figure}

We intervene independently on each level of the Conscientiousness personality trait in GPT-4o
and measure its response to the IPIP-NEO-300 personality inventory.
Figure~\ref{fig:induction-level} show the target level and the resulting inventory scores for
each personality trait.
The results show that all personality traits are relatively invariant
to changes in Conscientiousness.
Except for Neuroticism which is strong inversely correlated.

\subsection{Influence of personality trait level on CPT parameters}\label{subsec:influence-of-personality-trait-level-on-cpt-parameters}

\begin{figure}[!ht]
  \centering
  \includegraphics[width=\textwidth]{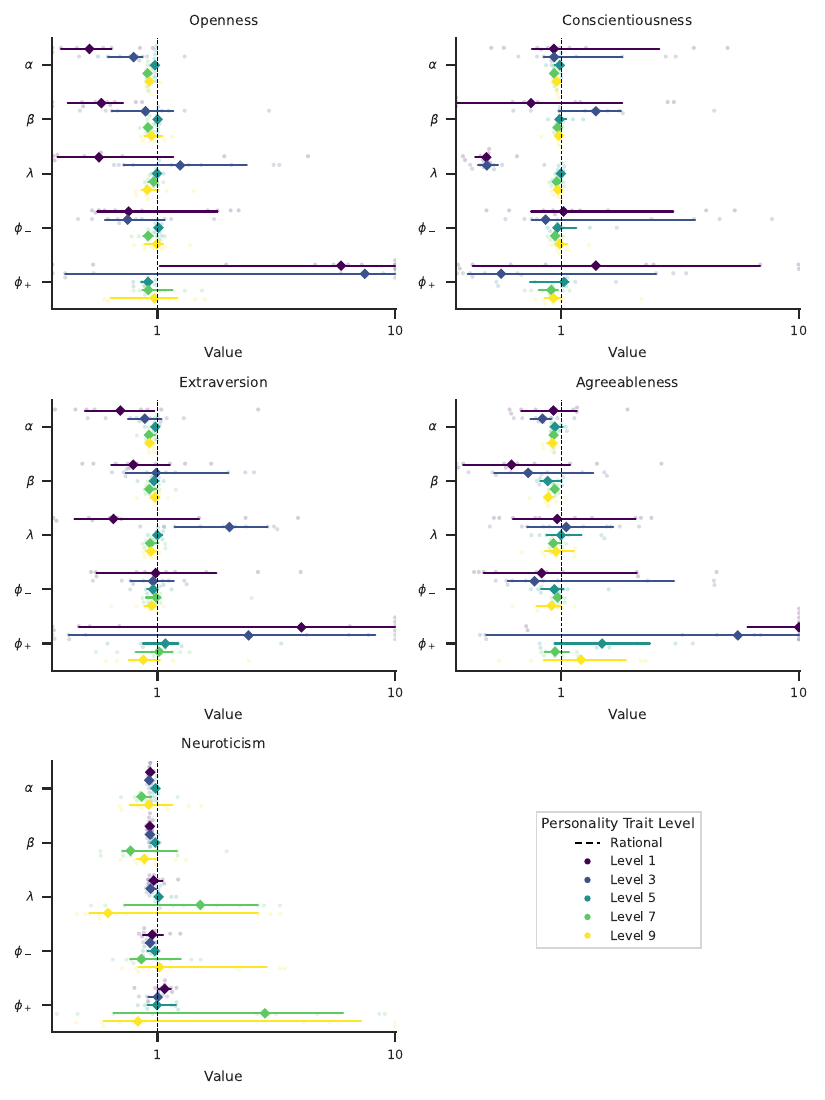}
  \caption{CPT parameter estimates for GPT-4o with personality interventions:
  The Big Five personality traits are induced at five levels, ranging from low to high.
  CPT parameters are estimated across $D_B$ over 10 runs, each with a unique random seed.
  Diamond markers represent the median values for each parameter,
  while error bars indicate the bootstrapped 95\% confidence interval.}
  \label{fig:personality-cpt-estimates}
\end{figure}

\begin{figure}
  \centering
  \includegraphics[width=\textwidth]{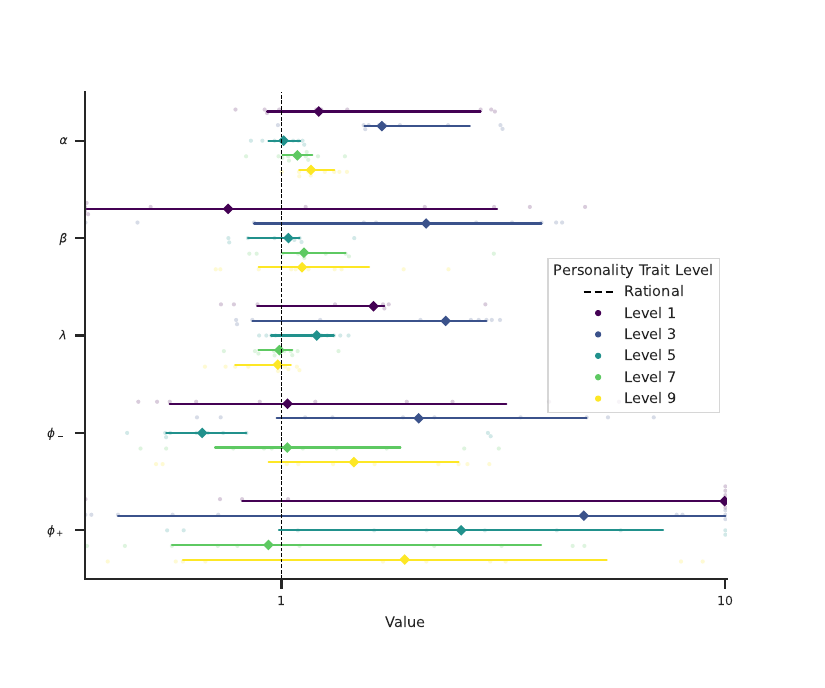}
  \caption{CPT parameter estimates for GPT-4-Turbo with personality interventions on Openness:
  The Big Five personality traits are induced at five levels, ranging from low to high.
  CPT parameters are estimated across $D_B$ over 10 runs, each with a unique random seed.
  Diamond markers represent the median values for each parameter,
  while error bars indicate the bootstrapped 95\% confidence interval.}
  \label{fig:gpt-4-cpt-openness}
\end{figure}

Figure~\ref{fig:personality-cpt-estimates} shows the CPT parameter estimates
for interventions on a personality traits in GPT-4o.
The results shows that risk-propensity is strongly influenced by Openness.
However the Neuroticism does not influence risk-propensity.
Figure~\ref{fig:gpt-4-cpt-openness} shows the CPT parameter estimates
for interventions on the personality trait Openness.
The results show localised patterns for levels 1--3 and levels 5--9.
This indicates that GPT-4-Turbo does not possess a global mapping from Openness to risk-propensity.

\subsection{System prompt ablations}
\label{subsec:comparisons}

\begin{figure}[!htb]
  \centering
  \begin{subfigure}[t]{0.45\textwidth}
      \centering
      \includegraphics[width=\textwidth]{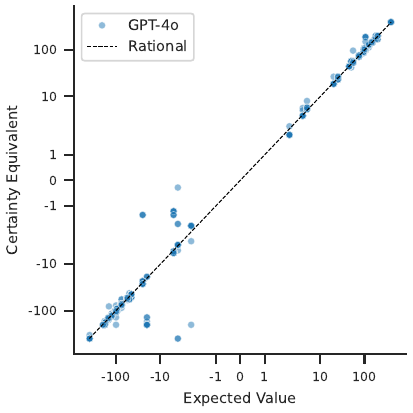}
      \subcaption{A comparison of the certainty equivalents for GPT-4o and the expected values of prospects in the dataset $D_A$ using the system prompt from \citet{ross2024llm}.}
      \label{fig:ce-ev-previous}
  \end{subfigure}
  \hfill
  \begin{subfigure}[t]{0.45\textwidth}
    \centering
    \includegraphics[width=\textwidth]{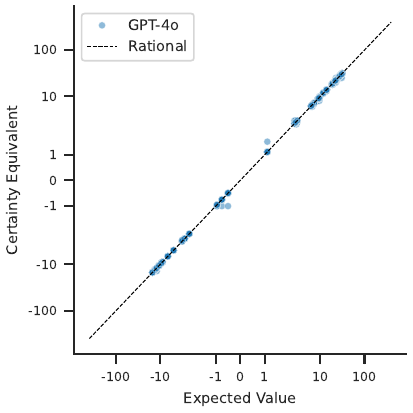}
    \subcaption{A comparison of the certainty equivalents for GPT-4o and the expected values of prospects in the dataset $D_A$ using the system prompt from our work.}
    \label{fig:ce-ev-our-work}
  \end{subfigure}
  \caption{An ablation of certainty equivalents and expected values for GPT-4o calculated using different methods.}
  \label{fig:sys-prompt-ablations}
\end{figure}

In this section we show the difference in certainty equivalents estimated using our
method and \citet{ross2024llm}.
Figure~\ref{fig:sys-prompt-ablations} shows the
estimated certainty equivalents over 15 runs.
Results for previous work show that certainty equivalents for prospects with potential losses are randomly distributed around positive and negative values.
Our method predicts more stable results since the certainty equivalents predicted by our method are closer to the expected value of the prospect (greater accuracy) and are more tightly clustered (greater precision).

\section{Additional information and datasets}\label{sec:additional-information-and-datasets}

\subsection{Model versions}\label{subsec:model-versions}

For GPT-4o we use \textit{gpt-4o-2024-05-13} and for GPT-4-Turbo we use \textit{gpt4-turbo-2024-04-09}.

\subsection{Big Five personality traits}
\label{subsec:big-five}

Table~\ref{tab:big_five_traits} shows personality facets in the IPI-NEO-300 inventory associated with each personality trait from the Big Five~\citep{goldberg1999broad}.

\begin{table}[!ht]
  \centering
  \caption{IPIP-NEO-300 facets and their associated Big Five Personality Traits.}
  \label{tab:big_five_traits}
  \begin{tabular}{p{0.3\textwidth}p{0.6\textwidth}}
  \toprule
  \textbf{Personality Trait} & \textbf{Facets} \\
  \midrule
  Openness& \makecell[l]{Fantasy, Aesthetics, Feelings,\\ Actions, Ideas, Values} \\
  \addlinespace
  Conscientiousness & \makecell[l]{Competence, Order, Dutifulness,\\ Achievement Striving, Self-Discipline,\\ Deliberation} \\
  \addlinespace
  Extraversion & \makecell[l]{Warmth, Gregariousness, Assertiveness,\\ Activity, Excitement-Seeking,\\ Positive-Emotions} \\
  \addlinespace
  Agreeableness & \makecell[l]{Trust, Straightforwardness, Altruism,\\ Compliance, Modesty,\\ Tender-Mindedness} \\
  \addlinespace
  Neuroticism & \makecell[l]{Anxiety, Angry Hostility, Depression,\\ Self-Consciousness, Impulsiveness,\\ Vulnerability} \\
  \bottomrule
  \end{tabular}
  \vspace{0.1cm}

  \end{table}

Table~\ref{tab:low-high-markers} shows the markers associated with each personality trait.
These markers are utilized in the system prompt to intervene on the language model's personality traits.

\begin{table}[!ht]
  \centering
  \caption{Low and high markers for the Big Five personality traits \citep{serapio2023personality}.}
  \label{tab:low-high-markers}
  \vspace{0.1cm}
  \begin{tabular}{|c|c|c|c|}
    \toprule
    \textbf{Trait} & \textbf{Low Level Marker (levels 1--3)} & \textbf{High Level Marker (levels 5--9)} \\
    \midrule
    Conscientiousness & unsure & self-efficacious \\
    Conscientiousness & messy & orderly \\
    Conscientiousness & irresponsible & responsible \\
    Conscientiousness & lazy & hardworking \\
    Conscientiousness & undisciplined & self-disciplined \\
    Conscientiousness & impractical & practical \\
    Conscientiousness & extravagant & thrifty \\
    Conscientiousness & disorganised & organized \\
    Conscientiousness & negligent & conscientious \\
    Conscientiousness & careless & thorough \\
    Extraversion & unfriendly & friendly \\
    Extraversion & introverted & extraverted \\
    Extraversion & silent & talkative \\
    Extraversion & timid & bold \\
    Extraversion & unassertive & assertive \\
    Extraversion & inactive & active \\
    Extraversion & unenergetic & energetic \\
    Extraversion & unadventurous & adventurous and daring \\
    Extraversion & gloomy & cheerful \\
    Agreeableness & distrustful & trustful \\
    Agreeableness & immoral & moral \\
    Agreeableness & dishonest & honest \\
    Agreeableness & unkind & kind \\
    Agreeableness & stingy & generous \\
    Agreeableness & unaltruistic & altruistic \\
    Agreeableness & uncooperative & cooperative \\
    Agreeableness & self-important & humble \\
    Agreeableness & unsympathetic & sympathetic \\
    Agreeableness & selfish & unselfish \\
    Agreeableness & disagreeable & agreeable \\
    Neuroticism & relaxed & tense \\
    Neuroticism & at ease & nervous \\
    Neuroticism & easygoing & anxious \\
    Neuroticism & calm & angry \\
    Neuroticism & patient & irritable \\
    Neuroticism & happy & depressed \\
    Neuroticism & unselfconscious & self-conscious \\
    Neuroticism & level-headed & impulsive \\
    Neuroticism & contented & discontented \\
    Neuroticism & emotionally stable & emotionally unstable \\
    Openness & unimaginative & imaginative \\
    Openness & uncreative & creative \\
    Openness & artistically unappreciative & artistically appreciative \\
    Openness & unaesthetic & aesthetic \\
    Openness & unreflective & reflective \\
    Openness & emotionally closed & emotionally aware \\
    Openness & uninquisitive & curious \\
    Openness & predictable & spontaneous \\
    Openness & unintelligent & intelligent \\
    Openness & unanalytical & analytical \\
    Openness & unsophisticated & sophisticated \\
    Openness & socially conservative & socially progressive \\
    \bottomrule
    \end{tabular}
\end{table}

Figure~\ref{fig:qualifiers} illustrates the mapping from markers in Table~\ref{tab:low-high-markers} to personality trait levels.
All markers for a given personality trait level are appended to each other for the intervention prompt.
For example, \textit{I'm very unsure, very messy, very irresponsible, very lazy, very undisciplined, very impractical, very extravagant, very disorganised, very negligent, very careless.}
is the intervention for Conscientiousness at level 2.

\lstset{
    language=Python,
    basicstyle=\ttfamily\small,
    breaklines=true,
    keywordstyle=\color{blue},
    stringstyle=\color{red},
    commentstyle=\color{green!60!black},
    numbers=left,
    numberstyle=\tiny\color{gray},
    numbersep=5pt,
    backgroundcolor=\color{gray!10},
    frame=single,
    framesep=5pt,
    rulecolor=\color{black!30},
    tabsize=4,
    showstringspaces=false
}

\begin{figure}[!htb]
  \begin{lstlisting}

  levels = {
    1: "extremely {low_marker}",
    2: "very {low_marker}",
    3: "{low_marker}",
    4: "a bit {low_marker}",
    5: "neither {low_marker} nor {high_marker}",
    6: "a bit {high_marker}",
    7: "{high_marker}",
    8: "very {high_marker}",
    9: "extremely {high_marker}",
  }
  \end{lstlisting}
  \caption{Pseudocode showing a mapping from personality prompt level to facet markers.
  Markers are give in Table~\ref{tab:low-high-markers}.}
  \label{fig:qualifiers}
\end{figure}

\subsection{Datasets}
\label{subsec:prospects}

In this section we provide a comprehensive description of the datasets used in our work.

\begin{figure}[!ht]
  \centering
  \begin{subfigure}[t]{0.43\textwidth}
      \centering
      \includegraphics[width=\textwidth]{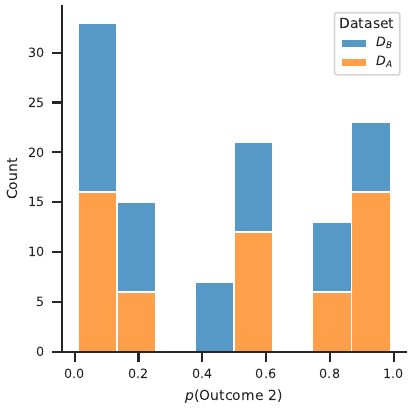}
      \subcaption{The distribution of probabilities associated with Outcome 2 for prospects in datasets $D_A$, $D_B$}
      \label{fig:dataset-probability-distribution}
  \end{subfigure}
  \hfill
  \begin{subfigure}[t]{0.43\textwidth}
    \centering
    \includegraphics[width=\textwidth]{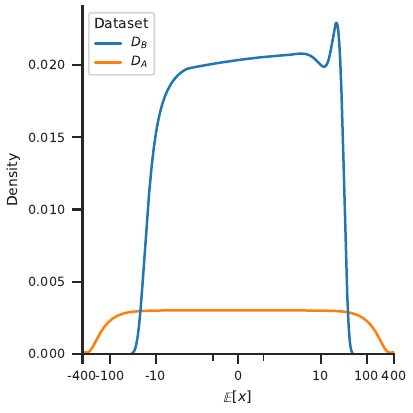}
    \subcaption{The distribution expected values for prospects in datasets $D_A$, $D_B$}
    \label{fig:dataset-expected-value-distribution}
  \end{subfigure}
  \caption{Summary statistics for the datasets $D_A$ and $D_B$}
  \label{fig:dataset-stats-summary}
\end{figure}

Figure~\ref{fig:dataset-stats-summary} shows the summary statistics for the datasets $D_A$ and $D_B$.

  \begin{table}[!ht]
    \centering
    \caption{Prospects in the dataset $D_A$ sampled from \citet{tversky1992advances}. \textrm{CE} is the median certainty equivalent estimated from 25 graduate students.
    The expected value of the prospect is $\mathbb{E}[x]$.}
    \label{tab:da-prospects}

    \vspace{0.1cm}
    \resizebox{0.5 \textwidth}{!}{
    \begin{tabular}{ccccc}
      \toprule
      \textbf{Outcome 1} & \textbf{Outcome 2} & \textbf{$p$} & \textbf{$\mathbb{E}[x]$} & \textbf{CE} \\
      \midrule
      0.00 & 50.00 & 0.10 & 5.00 & 9.00 \\
      0.00 & 50.00 & 0.50 & 25.00 & 21.00 \\
      0.00 & 50.00 & 0.90 & 45.00 & 37.00 \\
      0.00 & -50.00 & 0.10 & -5.00 & -8.00 \\
      0.00 & -50.00 & 0.50 & -25.00 & -21.00 \\
      0.00 & -50.00 & 0.90 & -45.00 & -39.00 \\
      0.00 & 100.00 & 0.05 & 5.00 & 14.00 \\
      0.00 & 100.00 & 0.25 & 25.00 & 25.00 \\
      0.00 & 100.00 & 0.50 & 50.00 & 36.00 \\
      0.00 & 100.00 & 0.75 & 75.00 & 52.00 \\
      0.00 & 100.00 & 0.95 & 95.00 & 78.00 \\
      0.00 & -100.00 & 0.05 & -5.00 & -8.00 \\
      0.00 & -100.00 & 0.25 & -25.00 & -23.50 \\
      0.00 & -100.00 & 0.50 & -50.00 & -42.00 \\
      0.00 & -100.00 & 0.75 & -75.00 & -63.00 \\
      0.00 & -100.00 & 0.95 & -95.00 & -84.00 \\
      0.00 & 200.00 & 0.01 & 2.00 & 10.00 \\
      0.00 & 200.00 & 0.10 & 20.00 & 20.00 \\
      0.00 & 200.00 & 0.50 & 100.00 & 76.00 \\
      0.00 & 200.00 & 0.90 & 180.00 & 131.00 \\
      0.00 & 200.00 & 0.99 & 198.00 & 188.00 \\
      0.00 & -200.00 & 0.01 & -2.00 & -3.00 \\
      0.00 & -200.00 & 0.10 & -20.00 & -23.00 \\
      0.00 & -200.00 & 0.50 & -100.00 & -89.00 \\
      0.00 & -200.00 & 0.90 & -180.00 & -155.00 \\
      0.00 & -200.00 & 0.99 & -198.00 & -190.00 \\
      0.00 & 400.00 & 0.01 & 4.00 & 12.00 \\
      0.00 & 400.00 & 0.99 & 396.00 & 377.00 \\
      0.00 & -400.00 & 0.01 & -4.00 & -14.00 \\
      0.00 & -400.00 & 0.99 & -396.00 & -380.00 \\
      50.00 & 100.00 & 0.10 & 55.00 & 59.00 \\
      50.00 & 100.00 & 0.50 & 75.00 & 71.00 \\
      50.00 & 100.00 & 0.90 & 95.00 & 83.00 \\
      -50.00 & -100.00 & 0.10 & -55.00 & -59.00 \\
      -50.00 & -100.00 & 0.50 & -75.00 & -71.00 \\
      -50.00 & -100.00 & 0.90 & -95.00 & -85.00 \\
      50.00 & 150.00 & 0.05 & 55.00 & 64.00 \\
      50.00 & 150.00 & 0.25 & 75.00 & 72.50 \\
      50.00 & 150.00 & 0.50 & 100.00 & 86.00 \\
      50.00 & 150.00 & 0.75 & 125.00 & 102.00 \\
      50.00 & 150.00 & 0.95 & 145.00 & 128.00 \\
      -50.00 & -150.00 & 0.05 & -55.00 & -60.00 \\
      -50.00 & -150.00 & 0.25 & -75.00 & -71.00 \\
      -50.00 & -150.00 & 0.50 & -100.00 & -92.00 \\
      -50.00 & -150.00 & 0.75 & -125.00 & -113.00 \\
      -50.00 & -150.00 & 0.95 & -145.00 & -132.00 \\
      100.00 & 200.00 & 0.05 & 105.00 & 118.00 \\
      100.00 & 200.00 & 0.25 & 125.00 & 130.00 \\
      100.00 & 200.00 & 0.50 & 150.00 & 141.00 \\
      100.00 & 200.00 & 0.75 & 175.00 & 162.00 \\
      100.00 & 200.00 & 0.95 & 195.00 & 178.00 \\
      -100.00 & -200.00 & 0.05 & -105.00 & -112.00 \\
      -100.00 & -200.00 & 0.25 & -125.00 & -121.00 \\
      -100.00 & -200.00 & 0.50 & -150.00 & -142.00 \\
      -100.00 & -200.00 & 0.75 & -175.00 & -158.00 \\
      -100.00 & -200.00 & 0.95 & -195.00 & -179.00 \\
      \bottomrule
      \end{tabular}}
  \end{table}

\begin{table}[!ht]
  \centering
  \caption{Prospects in $D_B$ randomly sampled from choices13k \citep{peterson2021using} The expected value of the prospect is $\mathbb{E}[x]$.}
  \label{tab:db-prospects}
  \vspace{0.5cm}
  \resizebox{0.4 \textwidth}{!}{
    \begin{tabular}{|c|c|c|c|}
      \toprule
      \textbf{Outcome 1} & \textbf{Outcome 2} & \textbf{$p$} & \textbf{$\mathbb{E}[x]$} \\
      \midrule
      29.00 & 37.00 & 0.05 & 29.40 \\
      16.00 & 47.00 & 0.50 & 31.50 \\
      -34.00 & 107.00 & 0.40 & 22.40 \\
      24.00 & 34.00 & 0.10 & 25.00 \\
      27.00 & 72.00 & 0.01 & 27.45 \\
      16.00 & 48.00 & 0.10 & 19.20 \\
      -14.00 & 37.00 & 0.10 & -8.90 \\
      -19.00 & 0.00 & 0.95 & -0.95 \\
      -16.00 & 16.00 & 0.80 & 9.60 \\
      2.00 & 90.00 & 0.01 & 2.88 \\
      -14.00 & -3.00 & 0.05 & -13.45 \\
      -28.00 & 38.00 & 0.60 & 11.60 \\
      3.00 & 26.00 & 0.25 & 8.75 \\
      -2.00 & 3.00 & 0.25 & -0.75 \\
      -46.00 & 70.00 & 0.60 & 23.60 \\
      18.00 & 20.00 & 0.10 & 18.20 \\
      -23.00 & 24.00 & 0.99 & 23.53 \\
      -7.00 & 10.00 & 0.80 & 6.60 \\
      -5.00 & -5.00 & 0.01 & -5.00 \\
      -31.00 & 100.00 & 0.40 & 21.40 \\
      -21.00 & 36.00 & 0.25 & -6.75 \\
      1.00 & 86.00 & 0.10 & 9.50 \\
      0.00 & 17.00 & 0.80 & 13.60 \\
      5.00 & 32.00 & 0.75 & 25.25 \\
      -12.00 & 58.00 & 0.60 & 30.00 \\
      -9.00 & 15.00 & 0.50 & 3.00 \\
      -7.00 & 35.00 & 0.50 & 14.00 \\
      -28.00 & 35.00 & 0.40 & -2.80 \\
      -16.00 & 3.00 & 0.90 & 1.10 \\
      -2.00 & 90.00 & 0.25 & 21.00 \\
      -13.00 & 15.00 & 0.01 & -12.72 \\
      -10.00 & 53.00 & 0.20 & 2.60 \\
      -10.00 & 29.00 & 0.99 & 28.61 \\
      -37.00 & -8.00 & 0.90 & -10.90 \\
      22.00 & 78.00 & 0.01 & 22.56 \\
      18.00 & 24.00 & 0.10 & 18.60 \\
      -23.00 & 82.00 & 0.40 & 19.00 \\
      -29.00 & 5.00 & 0.50 & -12.00 \\
      -9.00 & 25.00 & 0.25 & -0.50 \\
      -14.00 & 45.00 & 0.20 & -2.20 \\
      0.00 & 68.00 & 0.40 & 27.20 \\
      9.00 & 11.00 & 0.10 & 9.20 \\
      11.00 & 14.00 & 0.90 & 13.70 \\
      -15.00 & -4.00 & 0.75 & -6.75 \\
      -14.00 & 53.00 & 0.25 & 2.75 \\
      -9.00 & 9.00 & 0.90 & 7.20 \\
      22.00 & 30.00 & 0.10 & 22.80 \\
      -16.00 & 11.00 & 0.20 & -10.60 \\
      -24.00 & 28.00 & 0.40 & -3.20 \\
      -3.00 & 21.00 & 0.40 & 6.60 \\
      -44.00 & 2.00 & 0.75 & -9.50 \\
      -17.00 & 12.00 & 0.80 & 6.20 \\
      21.00 & 26.00 & 0.05 & 21.25 \\
      9.00 & 35.00 & 0.50 & 22.00 \\
      -9.00 & 70.00 & 0.50 & 30.50 \\
      -16.00 & 77.00 & 0.01 & -15.07 \\
      \bottomrule
    \end{tabular}
  }
\end{table}
\end{document}